\documentclass[aps,pre,longbibliography,showkeys,twocolumn,superscriptaddress]{revtex4-1}

\usepackage[english]{babel}
\usepackage{color}
\usepackage{listings}
\usepackage{float}
\usepackage{graphicx}
\usepackage{amsmath}
\usepackage{placeins}
\usepackage{hyperref}
\definecolor{dgreen}{rgb}{0,.4,0}
\definecolor{dblue}{rgb}{0,0,.7}
\usepackage{braket}
\usepackage{bbm}
\usepackage{seqsplit}
\usepackage{booktabs,array}

\begin{document}

\title{A Software Methodology for Compiling Quantum Programs}

\author{Thomas H\"aner}
\affiliation{Theoretische Physik, ETH Zurich, 8093 Zurich, Switzerland}
\author{Damian S. Steiger}
\affiliation{Theoretische Physik, ETH Zurich, 8093 Zurich, Switzerland}
\author{Krysta Svore}
\affiliation{Quantum Architectures and Computation Group, Microsoft Research, Redmond, WA (USA)}
\author{Matthias Troyer}
\affiliation{Theoretische Physik, ETH Zurich, 8093 Zurich, Switzerland}\affiliation{Quantum Architectures and Computation Group, Microsoft Research, Redmond, WA (USA)}\affiliation{Microsoft Research Station Q, Santa Barbara, CA (USA)}

\begin{abstract}
Quantum computers promise to transform our notions of computation by offering a completely new paradigm.
To achieve scalable quantum computation, optimizing compilers and a corresponding software design flow will be essential.
We present a software architecture for compiling quantum programs from a high-level language program to hardware-specific instructions.
We describe the necessary layers of abstraction and their differences and similarities to classical layers of a computer-aided design flow.
For each layer of the stack, we discuss the underlying methods for compilation and optimization.
Our software methodology facilitates more rapid innovation among quantum algorithm designers, quantum hardware engineers, and experimentalists. It enables scalable compilation of complex quantum algorithms and can be targeted to any specific quantum hardware implementation.
\end{abstract}


\maketitle

\section{Introduction}

The field of high-performance computing will be revolutionized by the introduction of scalable quantum computers.
Today, the majority of high-performance computing time is dedicated to solving problems in quantum chemistry and materials science. These problems would dramatically benefit from better classical algorithms or new models of computation that further reduce the processing time.
One such model, as suggested by Richard Feynman~\cite{Feynman1982}, is {\it quantum computation}, which takes advantage of quantum mechanics to obtain exponential speedups or improvements of solution.  Feynman proposed to use quantum computing to efficiently perform the simulation of physical systems~\cite{somma2002simulating}, precisely in areas such as quantum chemistry and materials science.
Since then, a variety of quantum algorithms have emerged, providing up to exponential speedups over their classical counterparts for problems in quantum chemistry~\cite{aspuru2005simulated,poulin2014trotter}, materials science~\cite{bauer2015hybrid}, cryptography~\cite{shor94}, and machine learning~\cite{lloyd2013quantum,rebentrost2014quantum,wiebe2014quantum,wiebe2014quantumn}. A complete list of algorithms can be found in Ref.~\cite{jordan2011quantum}.
In addition to quantum algorithmic improvements, significant advances in quantum hardware implementations have recently been made, suggesting imminent scalability.  The design of a scalable design flow for quantum computing is timely in order to program and compile to the rapidly evolving landscape of quantum devices.

\begin{figure}[tb]
\centering
\includegraphics[width=\linewidth]{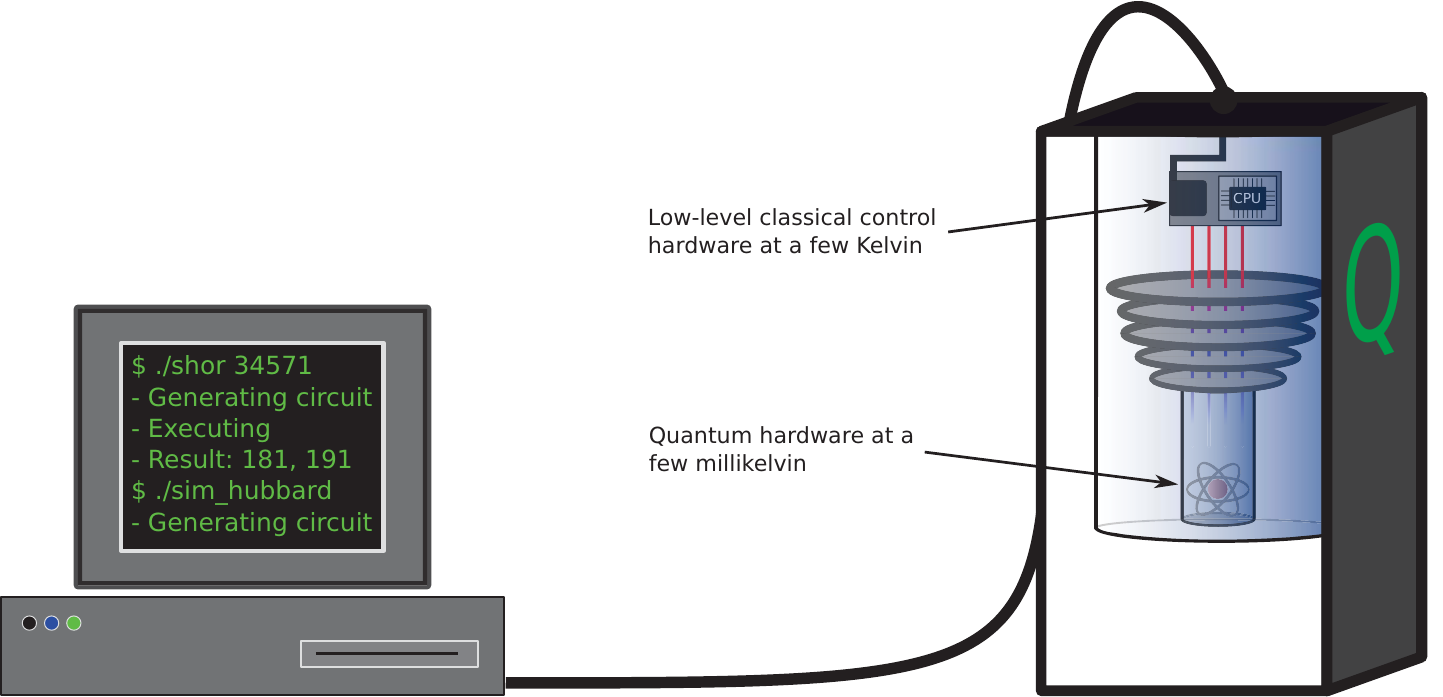}
\caption{Quantum computers will be used as accelerators controlled by classical hardware.
In many implementations, qubits must be kept at very low temperatures inside a cryogenic device.
Control of the system requires minimization of heat dissipation and fast communication, potentially requiring staging classical control processors and memory units at various temperatures, including at least one classical computer at room temperature.}
\label{fig:qcomp}
\end{figure}

Quantum computers are unlike existing hardware architectures.
A quantum computer will be intimately connected to a (potentially large) classical computer and will operate akin to a coprocessor, similar to GPUs today. See Figure \ref{fig:qcomp} for an illustration.
A classical computer will control the operations performed on the quantum computer and will also provide methods for maintaining a fault-tolerant computation.
The hardware architecture will require fast feedback between the quantum chip and the classical processor to ensure fault tolerance and a high quantum clock speed.
In many implementations, the quantum bits of information, called {\it qubits}, must be maintained at ultra-low temperatures inside a cryogenic device. In turn, classical processing and memory units for controlling the quantum computation will need to be located at several different temperatures, in order to minimize heat dissipation, maximize communication speeds between the classical and quantum hardware, and ensure minimal disturbance to the quantum system.

While classical computing benefits from a plethora of high-level programming languages and optimizing compilers, quantum computing remains mostly described at the level of logic operations, compiled and manually optimized. Recently, a handful of languages and compilers have been introduced. However, a complete end-to-end software methodolgy for compiling and optimizing quantum programs is lacking.
We propose a scalable software design flow for compilation and optimization of quantum programs.
By scalable, we mean the design enables programming any quantum algorithm of any size for any target architecture.

\begin{figure*}[t]
\centering
\includegraphics[width=\textwidth]{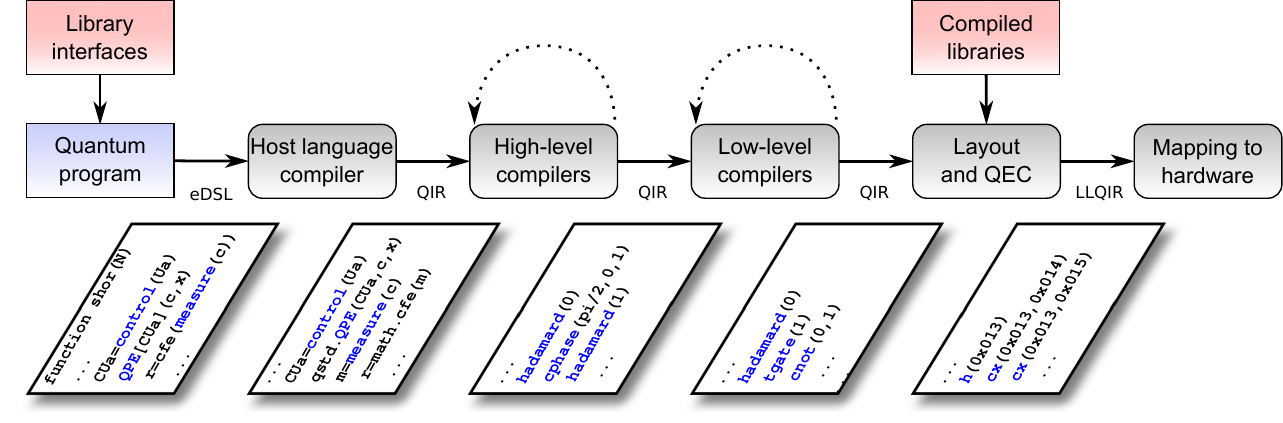}
\caption{The proposed toolchain for quantum programming: The quantum program is implemented in an embedded domain-specific language (eDSL) and then translated to hardware instructions using a series of compilers, which generate quantum intermediate representations (QIR) of the code. Pre-compiled libraries are linked against prior to creating the logical layout and applying error correction, since this will produce more efficient code. The resulting low-level quantum intermediate representation (LLQIR) is then mapped to hardware.}
\label{fig:toolchain}
\end{figure*}

Our software design flow is shown in Figure \ref{fig:toolchain}.
At the top of our design stack is a high-level language in which one writes quantum programs.
Quantum programs consist of an intimate mixture of classical and quantum instructions. Therefore, we argue the high-level language should be  an embedded domain specific language (eDSL) to maximally leverage existing classical infrastructure.
At the bottom, our methodology supports a vast variety of backends, such as simulators, emulators, resource analyzers, or any proposed quantum hardware implementation.
Our layered methodology provides modularity and flexibility for the future.
The high-level compilers are hardware-agnostic and, combined with quantum meta functions (e.g., conditional instructions and user annotations), enable an unprecedented level of optimization of quantum code.

Our software methodology enables rapid innovation in both quantum algorithms and hardware components.  
Its development in advance of a fully scalable hardware architecture allows verification of the software stack and testing of hardware constituents, through simulation and analysis of components.
It also enables the study of early design decisions, which for emerging technologies can result in substantial cost savings.
A scalable quantum software stack and hardware architecture is a complex system; our software design enables communication between algorithm designers, quantum engineers, and experimental physicists, and early definition of the necessary interfaces between the software and hardware systems. 
We outline a framework for automated programming of and compilation to a scalable quantum computer that is critical for early definition and development of software and hardware components, whether the hardware backend is an ion trap, superconducting system, quantum dot, or topological quantum computer.
\\\\\textbf{\textit{Related Work.}}
One of the earliest proposals for a scalable software methodology for quantum compiling dates back a decade \cite{svore2006layered} and presents significant steps of any quantum computing design flow.
Our work expands upon this work, extending the stack elements by showing how to integrate, e.g., tuned quantum libraries of arithmetic, subroutines, and quantum gates, and providing concrete details of the compilers and optimizers in the stack.
Later work presents a more detailed software architecture \cite{jones2012layered}, however the components are specific to a quantum dot architecture and the focus is mainly on the pipelined control cycle from the application layer down to the physical hardware layer.

High-level quantum programming languages have been more readily studied than supporting compilation frameworks \cite{gay2006quantum}.
Programming languages have been introduced that are based on C \cite{omer1998procedural} and based on functional languages \cite{green13,wecker14}.
We argue that instead of inventing a new programming language, a quantum software architecture should build on an embedded domain-specific language in order to leverage classical language and compilation features.

Existing quantum embedded languages include Quipper~\cite{green13} and LIQ$Ui\ket{} $\cite{wecker14}.
Both are functional languages and include an underlying compilation framework, allowing the representation of quantum algorithms and circuits at various levels of abstraction.  
Quipper is embedded in Haskell and allows extensible data types and advanced programming constructs.
 LIQ$Ui\ket{} $  contains a domain-specific language embedded in F\# ~\cite{wecker14}.
Compilation is primarily targeted to backends such as simulators and resource analyzers.
In contrast to Quipper,  LIQ$Ui\ket{} $ also features simulators for noise and resource analysis, but neither allows for hardware backends.

ScaffCC is a concrete implementation of a quantum compiler for a C-style language \cite{scaffcc14}. 
It compiles to a specific gate set in the QASM assembly language and enables program analysis and low-level optimizations.
Our methodology differs from the framework followed by ScaffCC in that it begins at the top with a high-level embedded language instead of a newly defined language, to enable easy programming of quantum algorithms and the ability to leverage existing classical language and compilation tools.  We also automatically translate to hardware instructions for specific backend quantum devices.
In addition, we introduce the compilation and auto-tuning of quantum libraries.  This allows programmers to reuse subroutines and automatically determine the best underlying circuit representations for the target backend, offering substantial quantum resource savings.

In addition to our demonstration implementations, our software methodology supports the Quipper and  LIQ$Ui\ket{} $ languages as the high-level programming language, and can leverage their simulation and resource analysis tools within various layers of the stack.
However, our design significantly extends and defines the necessary components of a complete scalable compilation framework, including details on how to compile, tune, and link libraries, where to map to an explicit hardware layout and error correction, and how to map to hardware controls and instructions.  We detail the  necessary layers of abstraction and the optimizing transformations that will ultimately lead to successful compilation of any quantum algorithm for quantum hardware.  We have designed our framework explicitly to program and control a quantum computer, no matter its size.
\\\\\textbf{\textit{Outline.}}
In Section \ref{sec:qc}, we briefly review quantum computation and highlight key characteristics that prove crucial in the design of a software architecture.
We introduce quantum programs in Section \ref{sec:qprograms} and define quantum types, quantum gates, quantum libraries, and quantum algorithms.
In Section \ref{sec:compilation}, we step through each compilation layer of the proposed software stack.  We begin by discussing the high-level compilers, including the host language compiler and high-level quantum compiler, and then describe the low-level compiler, including details on the challenges of fault tolerance and layout generation for quantum computers.
In Section \ref{sec:backends}, we describe the various backends that can be targeted.  We highlight the value of auto-tuning of quantum libraries and optimization of representations in Section \ref{sec:autotuning} and close with a discussion of design implementations of our stack and future work in Sections \ref{sec:impl} and \ref{sec:summary}.

\section{Quantum Computing}\label{sec:qc}
Quantum computers harness quantum effects in order to speed up calculations. Information is stored in quantum bits called qubits and computations are performed by applying quantum gates and measurements to the quantum state of the qubits.

In contrast to its classical counterpart, a qubit may be in an arbitrary superposition of its two basis states labeled $\Ket0$ and $\Ket1$:
\begin{equation}\label{eq:superpos}
	\alpha\Ket0+\beta\Ket1\;,
\end{equation}
with complex amplitudes $\alpha,\beta\in\mathbbm C$ satisfying the normalization condition $|\alpha|^2+|\beta|^2=1$.
 The state of a general $n$-qubit system can be an arbitrary superposition over all $2^n$ computational basis states, i.e.,
 \begin{equation}
 \nonumber
 \sum_{q_1q_2...q_n\in \{0,1\}^n} c_{q_1...q_n}\Ket{ q_1...q_n}   = \sum_{i=0}^{2^n-1}c_i\Ket i\;,
 \end{equation}
where we have interpreted the basis state $q_1...q_n$ as a binary number and replaced it with the corresponding integer $i$. Again, the complex amplitudes $c_i$ need to satisfy the normalization condition $\sum_i |c_i|^2=1$.
 
Information stored in qubits is retrieved by measurements, which convert qubits into classical bits. In the case of measuring all qubits simultaneously, the outcome is one of the basis states with probability equal the absolute value squared of the complex amplitude $c_i$ of that basis state $i$.
After the measurement, the qubits are no longer in superposition but in the state determined by the measurement. When measuring a single qubit in the superposition state \eqref{eq:superpos}, the outcome is either 0 or 1 with probability $|\alpha|^2$ or $|\beta|^2$, respectively, and the wavefunction of the qubit collapses onto the respective basis state ($\ket0$ or $\ket1$).

Similar to classical operations, quantum gates can be applied to qubits to change their state. Multi-qubit gates are very hard to realize in hardware. However, there exist universal gate sets consisting of only a handful of one- and two-qubit gates, that is, gates applied to just one or two qubits simultaneously, which can be used to implement any $n$-qubit quantum gate.

As a consequence of the linearity of quantum mechanics, gates are simultaneously applied to all basis states of a superposition at once -- similar to classical SIMD operations. If, for example, $n$ qubits are in a complete superposition over all $2^n$ basis states, all possible outputs of a function $f(x)$ can be calculated using only one function call:
\begin{equation*}
f(\sum_{i=0}^{2^n-1} c_i \ket{x}) = \sum_{i=0}^{2^n-1}  c_i \ket{f(x)}\;.
\end{equation*}
This phenomenon is called quantum parallelism. However, the outputs of all $2^n$ computations are saved in a quantum superposition. Therefore, simply measuring the qubits will just pick one result at random and is thus no more powerful than classical computation.  To overcome this, quantum algorithms need to employ sophisticated reduction schemes, making use of quantum interference effects. For a more thorough introduction to quantum computing and a description of the most common gates, see the Appendix.

Throughout this paper, we will illustrate quantum circuits through diagrams, where each wire represents a qubit and boxes represent quantum gates.  Similar to classical circuit diagrams, these quantum circuit diagrams are read from left to right, such that gates at the left of the circuit are applied before those to the right.

\section{Quantum programs}\label{sec:qprograms}

\begin{figure}[t]
	\centering
	\includegraphics[width=.8\linewidth]{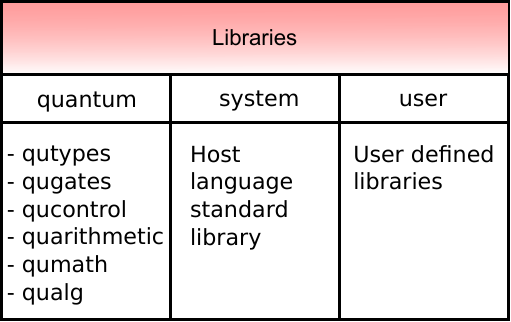}
	\caption{Libraries available to the quantum program. The quantum standard library is described in the text. The system (or standard) library is provided by the host language. The user may add custom quantum or classical libraries in order to increase code reuse among developers.}
	\label{fig:libraries}
\end{figure}

A quantum program consists of a sequence of classical and quantum instructions to be executed on hybrid systems consisting of both classical and quantum hardware. To lift the quantum parts of the codes to the same level of abstraction as classical codes, we envision three types of libraries that can be leveraged to write quantum/classical hybrid programs using an embedded DSL (see Figure~\ref{fig:libraries}): The classical standard or system library (from the host language), custom user libraries, and the quantum libraries that we describe below.
\\\\\textbf{\textit{Quantum types.}}
	In classical computing, one prefers to use abstract types instead of operating at bit-level. Therefore, such types need to be available for quantum programming as well. In particular we propose:
	\begin{itemize}
	\item\texttt{qubit}: A quantum bit, which is the basic building block of all quantum types
         \item\texttt{qureg}: General purpose quantum registers, consisting of a fixed or variable number of qubits	
		\item\texttt{quint}: Quantum integer types
		\item\texttt{qufixed}: Quantum fixed-point types
		\item\texttt{qufloat}: Quantum floating-point types
	\end{itemize}
Given the extremely limited number of qubits in early quantum computers, it is important to provide versions of \texttt{qureg}, \texttt{quint}, \texttt{qufixed} and \texttt{qufloat} with variable numbers of bits. The numerical types also provide functions and operator overloads, which facilitate the use of the libraries discussed below. Being aware of the limited number of operations that early quantum computers will be able to perform, the use of fixed point arithmetic is crucial and will be more common than the use of floating point arithmetic. 
\\\\\textbf{\textit{Quantum gates.}}
	Quantum gates can act on one or multiple qubits. Similar to classical computers, there are a few standard gates which frequently occur in quantum algorithms. A list of common gates and their definitions is provided in the Appendix. 
	
	The quantum gate library features a variety of gates chosen based on various technologies and applications. Gates that are not supported by the hardware must be decomposed by the compiler.

\begin{figure}[!t]
	\centering
	\includegraphics[width=\linewidth]{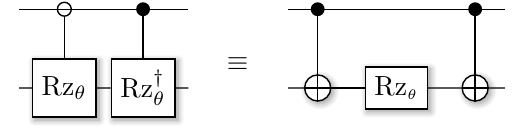}
	\caption{Providing a \texttt{quifelse} meta function allows the compiler to reduce the number of gates from ten to three in this example of a conditional inverse rotation.}
	\label{fig:cadj}
\end{figure}

	In contrast to classical gates, many quantum gates, such as rotation gates, have continuous parameters and cannot be executed to arbitrary precision in an actual device. Therefore, the programmer needs the option to override the conservative default tolerances for individual gates or for an entire algorithm. These tolerances are then used to optimize hardware control and gate synthesis.  Gate synthesis is the process by which any single-qubit quantum gate can be implemented to any desired precision by finding an efficient sequence of gates drawn from a discrete set of well-calibrated and error corrected gates.  Efficient methods for such gate synthesis are well studied \cite{bocharov2015efficient,ross2014optimal}. 
	\\\\\textbf{\textit{Quantum control flow statements.}}
Quantum algorithms often contain certain computational patterns that are not common in classical computing. Making these patterns explicit through the use of meta functions makes programming easier and facilitates aggressive code optimization. Among others, the following meta operations are provided:

	\begin{itemize}
			\begin{lstlisting}[float,floatplacement=H,caption={One possible way of introducing code annotation to help the compiler identify compute-action-uncompute sequences, which can be turned into controlled versions at a much lower cost.},label={lst:cau}]
compute(QFT(x)) // basis change
phiAdd(x,a)   // a+x in Fourier space
uncompute()   // applies inverse(QFT)
		\end{lstlisting}
		
		\item \texttt{compute/uncompute}: As quantum operations must be reversible, temporary variables can only be discarded after reversing their computation by an uncompute step~\cite{bennett73}.  That is, ancilla qubits (scratch space) must be clean at the end of the computation, or before reuse as scratch space. Making the compute/uncompute pattern explicit by annotating the compute section allows the uncomputation to be done automatically by the compiler. Furthermore, various high-level optimizations can be performed, as discussed below. In certain cases, the user may override the default uncomputation by an optimized implementation thereof. See Listing~\ref{lst:cau} for an example use case.	
	
	\begin{figure}[!t]
		\centering
		\includegraphics[width=\linewidth]{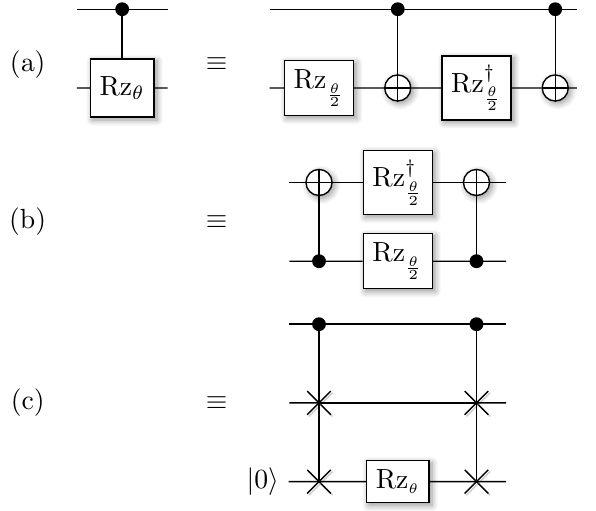}
		\caption{Three possible decompositions of a controlled z-rotation using single-qubit z-rotations and either CNOTs~\cite{barenco1995elementary,ito2012} or Fredkin gates (controlled swaps)~\cite{wiebe2012}. The first option features higher locality of rotations and the second allows for parallelization thereof. The third option requires only one rotation, at the cost of an extra qubit. Note that a single control already doubles the number of rotations and increases the number of CNOTs by 2 (for (a) and (b)) or requires two Fredkin gates and one additional qubit (for (c)). Therefore, the reduction of the number of controlled gates is essential for arriving at efficient quantum code.}
		\label{fig:crot}
	\end{figure}
	
		\item \texttt{control}: Conditional code is not straightforward on quantum computers since the condition variable can be invoked in a superposition of being both true and false. Similar to classical SIMD codes, one has to use masking and apply the operations only to those computational states where a control qubit is 1. 	
		This can be achieved by replacing every gate by its controlled version. However, this significantly increases the number of gates. For example, as shown in  Figure~\ref{fig:crot}(a), for simple rotation gates the number of gates is four times higher in the controlled version.
		Therefore, it is important to reduce the number of controlled operations by using high-level information such as the \texttt{compute/uncompute} pattern, where the compute and uncompute steps themselves do not need to be controlled, as depicted in Figure \ref{fig:cau}. The reason is that when the central operation $V$ is masked out by the control, then the uncompute steps just undoes the compute step and their combined action is trivial. It is thus not necessary to control these compute and uncompute steps, resulting in signifcant savings.
		
		\begin{figure}[!t]
			\centering
			\includegraphics[width=\linewidth]{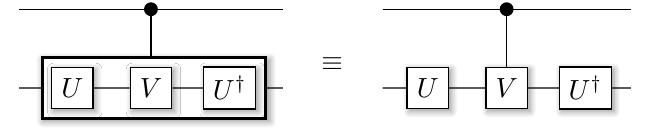}
			\caption{Controlled compute/action/uncompute-sequences can be achieved by only controlling the middle operation, thus significantly reducing the number of controlled gates.}
			\label{fig:cau}
		\end{figure}

		\item \texttt{repeat}: A straight-forward generalization of loops to the quantum domain is not possible due to the fact that the resulting circuit must be able to handle a superposition of inputs. Therefore, if the loop condition is quantum, an upper-bound on the number of iterations must be provided. Furthermore, the entire loop body must be conditioned on a continuation indicator qubit in order to handle cases where the loop would terminate before the maximal number of iterations is reached.
		
		If, on the other hand,  the loop condition only depends on classical information, such as the outcome of measurements, the handling is simpler. In such cases, the code can be optimized by sending the entire loop instructions to lower-level classical hardware, thereby greatly reducing the overhead of communication between the host and the quantum device.
		
		\item \texttt{quifelse}: This meta function is a generalization of the control operation, introduced in order to catch optimization opportunities, where one operation is performed if a control qubit is 0 and another operation is performed if it is 1. One common example is the conditional inverse for rotating in one direction if the control is 0 and the opposite direction if it is 1. Instead of performing two controlled rotations, which would lead to a total of ten gates (see (a) and (b) in Figure~\ref{fig:crot}), this is optimized as shown in Figure~\ref{fig:cadj}, resulting in a reduction of the number of rotations by a factor of four ~\cite{wecker15}. 
	\end{itemize}	
\textbf{\textit{Quantum arithmetic.}}
	The quantum arithmetic libraries feature optimized low-level implementations of basic arithmetic functions such as addition, subtraction, multiplication, fused multiply-add, division, and inverses, which can then be used to implement high-level functions. 
	Upon compilation, the best available implementation can be chosen for the quantum architecture in question using an auto-tuning approach similar to classical linear algebra packages~\cite{whaley98} -- see Section \ref{sec:autotuning}.
	\\\\\textbf{\textit{Quantum math library.}}
	Building on the quantum arithmetic library, the math library provides implementations of high-level mathematical functions such as exponential and arcsine. Again, depending on the quantum hardware being compiled to, different approaches may turn out to be most efficient. Finding the best one can be achieved using an auto-tuning approach which takes into account both the compilation of the underlying arithmetic operations and e.g., the type of series expansion used. In addition, there might be precision-dependent implementations, which is especially useful in this case, as qubits are currently a very limited resource.
	
	Implementing such a quantum math library is a highly non-trivial endeavor due to the reversibility condition of quantum algorithms. As an example, consider Newton's method, which can be used to implement inverses of trigonometric functions in classical computing. A straight-forward translation of this approach is not feasible in the quantum domain due to the fact that the method needs to succeed for a superposition of values. Therefore, the number of iterations would have to be chosen such that all values converge to the correct result and simply determining all outcomes scales exponentially with the number of qubits.
	
	Similar parallelism and the difficulty arising from it can also be observed in classical computing when dealing with vector units, which perform single instructions on multiple data (SIMD). Hence, these may serve as a starting point for efficient  implementations of mathematical functions on quantum hardware.
\\\\\textbf{\textit{Quantum algorithms.}}
	This highest level algorithms library provides standard quantum subroutines such as the quantum phase estimation (QPE), which estimates eigenvalues of a unitary matrix, the quantum Fourier transform (QFT), which performs a Fourier transformation on the wave function,  linear system solvers, and other common quantum algorithms. Similar to classical programming, the availability of  higher level algorithms accelerates quantum programming. Furthermore, the compilation process can pick implementations optimized for specific target hardware. Providing such high-level quantum algorithms will accelerate the development of new quantum programs.

\FloatBarrier
\section{Compilation process}\label{sec:compilation}
\begin{figure*}
\centering
\includegraphics[height=.9\textheight]{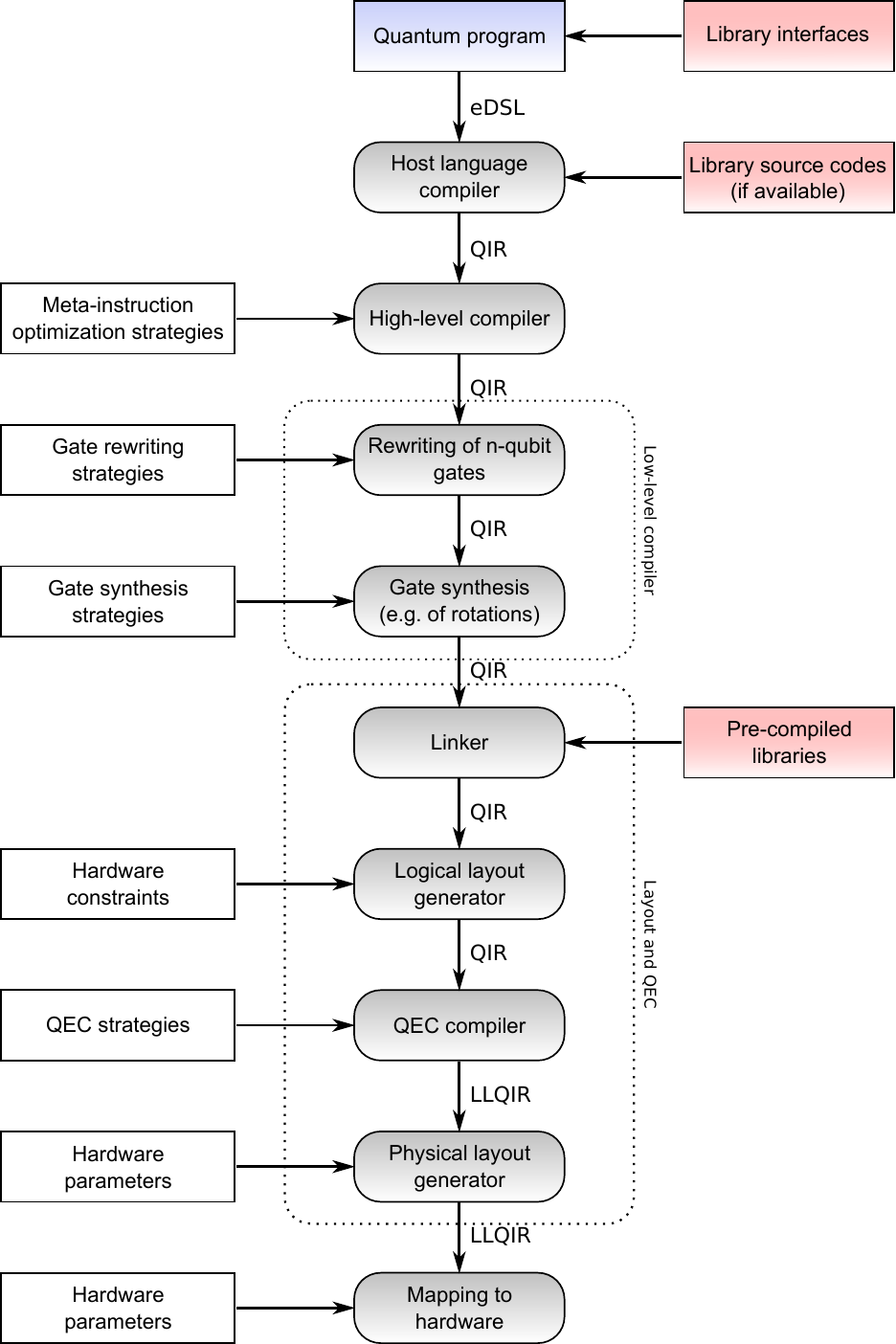}
\caption{Detailed view of the toolchain: The quantum program written in an embedded domain-specific language (eDSL) is translated to hardware instructions using a series of compilers, which generate quantum intermediate representations (QIR) and, ultimately, low-level quantum intermediate representations (LLQIR) of the code. As a final step, the LLQIR is translated to hardware instructions.}
\label{fig:detailedtoolchain}
\end{figure*}

The translation of a quantum program down to hardware instructions is a layered process consisting of several  compilation and optimization steps. Each compiler generates a quantum intermediate representation (QIR) or -- after error correction -- a low-level QIR (LLQIR) of the original code. The layering allows for a hardware-agnostic implementation of the higher-level compilers, thus increasing code reuse. The overview of the compilation process is depicted in Figure~\ref{fig:toolchain} and a more detailed version, whose individual steps are discussed in the following, is shown in Figure~\ref{fig:detailedtoolchain}.

\begin{figure}[!t]
	\centering
	\includegraphics[width=\linewidth]{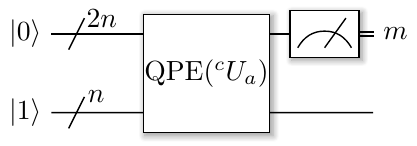}
	\caption{Circuit representation of Shor's algorithm. A quantum phase estimation (QPE) is performed using a controlled version of the operator $U_a$, which implements $U_a\Ket x=\Ket{ax\text{ mod }N}$. The result of the QPE can then be used to extract the period of the function $f(x)=a^x\text{ mod }N$ and, thus, determine the factors of $N$.}
	\label{fig:shorhl}
\end{figure}

The stages of compilation will be illustrated using a high-level implementation of Shor's algorithm for factoring \cite{shor94}. While we have implemented demonstration versions of this architecture and Shor's algorithm in a number of programming languages, some of which are discussed below, we explain the architecture using pseudocode to highlight the software requirements and not specific implementations.
Listing~\ref{lst:shorcodehl} shows pseudocode and Figure~\ref{fig:shorhl}  the corresponding circuit diagram for an implementation similar to that of Ref.~\cite{beauregard02}.  Note that the final box represents measurement.

\begin{lstlisting}[caption={High-level pseudocode of Shor's algorithm, which needs to be compiled down to a given specific hardware.},float,floatplacement=H,label=lst:shorcodehl]
...
x = allocateQureg(n, value = 1)
ancillas = allocateQureg(2n)
Ua(k,x) = a(k)*x mod N 
CUa(k,c,x) = control(c,Ua(k,x))
QPE[CUa] (ancillas, x)
m = measure(ancillas)
r = continuedFracExpand(m).denom()
...
\end{lstlisting}

\subsection{Host language compilation} The first step of the compilation process is carried out by the host language compiler/interpreter, resolving  the control statements of the classical code and performing first optimizations. It also dispatches quantum statements to (potentially backend-dependent) library functions. Having the classical compiler do as much as possible keeps the effort of implementing the compilation framework minimal.

\textit{Shor example:} Sending the high-level code in Listing~\ref{lst:shorcodehl} through this compiler results in code similar to what is shown in Listing~\ref{lst:shorcodehlhost}: In this pseudocode example, the resource allocation (\texttt{allocateQureg}) has been replaced by a member function call of the backend for which the code is being compiled. The backend may be anything that fulfills a predefined concept, be it an emulator or an object which sends hardware instructions to the quantum device. Furthermore, the mathematical expression has been resolved to a quantum math function call. Similar transformations have been applied to the QPE-syntax and measurement, which should be handled by a concrete backend as well.

\begin{lstlisting}[caption={High-level pseudocode of Shor's algorithm after going through the host language compiler.},float,floatplacement=H,label=lst:shorcodehlhost]
...
x = backend.allocateQureg(n)
backend.initialize(x, 1)
ancillas = backend.allocateQureg(2n)
Ua(k,x) = qmath.axmodN(a(k), x, N)
CUa(k,c,x) = control(c, Ua(k,x))
qstd.QPE(CUa, ancillas, x)
m = backend.measure(ancillas)
r = continuedFracExpand(m).denom()
...
\end{lstlisting}

\subsection{High-level quantum compiler} Following the host language compiler/interpreter, the quantum part of the code is compiled further using all available information on the high-level structure of the quantum program. This includes both meta instruction replacement and optimization.

\textit{Replacements:} Meta functions and library calls are replaced by their implementation, where available. For example, the \texttt{uncompute} statement applies the inverse of the previous \texttt{compute}-section.  Another example is the \texttt{control} instruction, which may be applied to individual gates or entire algorithms. Since libraries may provide controlled versions of implemented functions, the compiler chooses between either directly calling a controlled version of the function, annotating the library call with a control-flag that will be resolved in later stages, or compiling a controlled version of the function if the source code is available. In the latter case, the compiler makes use of all meta-information available (e.g., that the compute/uncompute sections do not need to be controlled) in order to make the resulting code as efficient as possible.

\textit{Optimizations:} After all meta functions have been handled, the compiler executes a series of optimization steps in order to improve code efficiency. For example, multiple rotations carried out on the same qubit can be combined in a quantum analog of constant folding, which would be much harder after rotation gates have been synthesized into fundamental gates further down the stack. 

\begin{figure}[!ht]
	\centering
	\includegraphics[width=\linewidth]{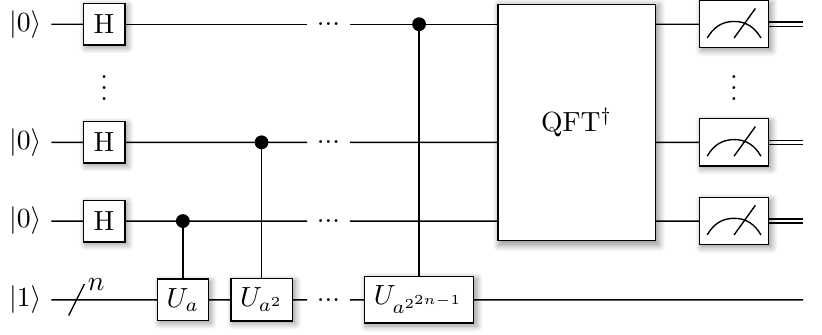}
	\caption{Shor circuit where the library implementation of QPE has been inlined. The inverse quantum Fourier transform (QFT$^\dagger$), which is applied to the $2n$ qubits at the top, is another library call yet to be handled by the compiler.}
	\label{fig:shorml}
\end{figure}

\textit{Shor example (continued):} Applying these first transformations to the high-level circuit of Shor's algorithm, which is depicted in Figure~\ref{fig:shorhl}, yields the circuit in Figure~\ref{fig:shorml} at first, where the library implementation of the quantum phase estimation has been inlined. If the user were to provide a quantum register whose size is incompatible with the desired precision, an iterative phase estimation would be chosen instead of the standard QPE.

\begin{figure}[!ht]
	\centering
	\includegraphics[width=\linewidth]{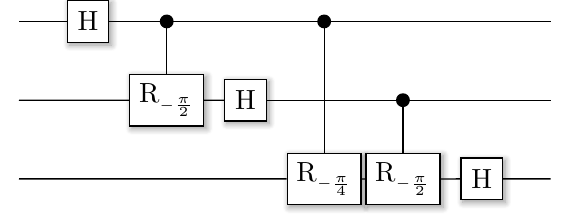}
	\caption{Possible library implementation of a 3-qubit inverse quantum Fourier transform.}
	\label{fig:qft}
\end{figure}

Now, the compiler proceeds by resolving the controls on the function $U_a$, making use of the structure of the computation (e.g., not controlling basis changes). In addition, it uses the library implementation of the inverse quantum Fourier transform (QFT$^\dagger$), which itself might require optimization. A possible library implementation for a 3-qubit QFT$^\dagger$ is depicted in Figure~\ref{fig:qft}. 

This entire process is iterated until there are no library calls left for which the source code is available. An excerpt of the quantum intermediate representation after having completed this stage is depicted in Listing~\ref{lst:shorcodeml}.

\begin{lstlisting}[caption={Excerpt of a quantum intermediate representation of Shor's algorithm after a few iterations of the high-level compilation.},float,floatplacement=T,label=lst:shorcodemlfirst]
...
/* QPE on modular multiplier.
   Do repeated modular addition and
   shift on x, store result into b */
// begin modular Draper-addition
ccphiadd(qpe_ctrl, x[i], a, b)
phisub(N, b)
iQFT(b)
backend.apply(CNOT, b[end], ancilla)
QFT(b)
cphiadd(ancilla, N, b)
ccphisub(qpe_ctrl, x[i], a, b)
iQFT(b)
backend.apply(NOT, b[end])
backend.apply(CNOT, b[end], ancilla)
backend.apply(NOT, b[end])
QFT(b)
ccphiadd(qpe_ctrl, x[i], a, b)
...
m = backend.measure(0, ..., 2n-1)
...
\end{lstlisting}


\subsection{Low-level quantum compiler} After the high-level compilation stage, the code consists of quantum gates, inlined library functions, and  library calls to be resolved later. The task of the low-level compiler is to translate all quantum gates into sequences of gates from a discrete, technology-dependent set of quantum gates. The concrete structure of this set depends on both the hardware and the chosen error-correction strategy.

\textit{Rewriting and synthesis of gates:} Each quantum gate now has to be synthesized from a smaller, hardware-specific set of gates. For example, multi-qubit gates will be synthesized from one-qubit gates and a technology-specific two-qubit gate, similar to what is depicted e.g. in Figures~\ref{fig:crot}(a) and \ref{fig:crot}(b). Then, further hardware-specific rewriting rules are applied, including the decomposition of single-qubit operations into sequences of gates from a technology-dependent set, which could include the so-called T, H, and S gates. This gate synthesis can be achieved deterministically~\cite{ross2014optimal} or using non-deterministic algorithms~\cite{bocharov2015efficient}. 

\begin{lstlisting}[caption={Excerpt of a quantum intermediate representation of Shor's algorithm after the final iteration of the high-level compilation.},float,floatplacement=T,label=lst:shorcodeml]
...
// begin QPE
...
// begin inverse QFT
backend.apply(H, 0)
backend.apply(CR(-pi/2), 0, 1)
backend.apply(H, 1)
backend.apply(CR(-pi/4), 0, 2)
...
m = backend.measure(0, ..., 2n-1)
...
\end{lstlisting}

\textit{Optimizations:}
As a final step of the low-level compilation phase, local optimization rules can be applied with the goal of further reducing the number of expensive operations, such as the number of T gates on certain architectures, see, e.g., Ref.~\cite{maslov2008quantum}.

\textit{Shor example (continued):} The rewriting and synthesis of gates can be observed in our Shor example, where this decomposition is applied to the rotation gates that arise from the inverse quantum Fourier transform. The conditional phase shift of $-\frac\pi2$ can be rewritten in terms of T, S, and CNOT gates, as shown in Listing~\ref{lst:shorcodell}. The rotations required for the $-\frac\pi4$ controlled phase shift, on the other hand, need to be approximated using one of the approaches mentioned above. Considering that the respective sequences for a rotation gate with an angle $\theta=-\frac{\pi}{8}$ is
\\
\seqsplit{HTSHTSHTSHTHTSHTSHTSHTHTSHTHTSHTHTSHTSHTHTHTSHTSHTHTSHTHTSHTHTHTHTSHTSHTSHTHTSHTSHTSHTHTSHTHTHTHTSHTHTHTSHTHTHTHTSHTSHTSHTHTSHTSHTHTSHTHTSHTHTSHTHTSHTSHTHTSHTHTSHTSHTSHTSHTHTHTSHTHTSHTHTHTHTSHTHTHTHTHTHTSHTSHTSHTHTSHTSHTHTSHTSHTHTHTSHTSHTHTHTSHTHTHSS} 
\\for a tolerance of  $10^{-10}$  and\\
\seqsplit{HTHTSHTSHTSHTSHTHTSHTHTHTSHTHTSHTHTHTSHTSHTSHTSHTSHTSHTSHTSHTSHTHTSHTHTSHTHTHTSHTSHTSHTHTHTHTSHTSHTHTSHX}
\\for a tolerance of $10^{-4}$, one immediately sees the trade-off between runtime and accuracy.

In the end, the quantum intermediate representation consists of code similar to the one shown in Listing~\ref{lst:shorcodell}.
\begin{lstlisting}[caption={Part of the quantum intermediate representation of Shor's algorithm after the low-level compilation. The controlled rotation gates have been decomposed into CNOT, S, H, and T gates.},float,floatplacement=T,label=lst:shorcodell]
...
// begin inverse QFT
backend.apply(H, 0)
// decompose Rz(-pi/4) (trivial)
backend.apply(T, 1)
backend.apply(S, 1)
backend.apply(S, 1)
backend.apply(S, 1)
backend.apply(CNOT, 0, 1)
backend.apply(T, 1)
backend.apply(CNOT, 0, 1)
backend.apply(Tdagger, 0)
backend.apply(H, 1)
// decompose Rz(-pi/8)
backend.apply(H, 2)
backend.apply(T, 2)
backend.apply(S, 2)
backend.apply(H, 2)
backend.apply(T, 2)
backend.apply(S, 2)
...
\end{lstlisting}

\subsection{Logical layout generator and optimizer}
Following the low-level compilers, we have to assign qubit variables to logical qubits similar to how variables are assigned to registers in classical computers.

The code coming from the low-level compilers still contains calls to library functions such as arithmetic operations. While a large-scale quantum computer might have dedicated qubits and gates for certain functions similar to the arithmetic and logical units in modern CPUs, early quantum computers will only have a small number of general purpose qubits and gates. Hence, calls to pre-compiled libraries now need to be replaced by the corresponding code,  which has been optimized for the specific hardware. At this point, the logical layout can be done.

So far we have only talked about qubits in an abstract way. We now need to introduce the notion of three different types of qubits: \emph{logical qubit}, \emph{physical qubit} and \emph{hardware qubit}.

\emph{Logical qubits} are qubits used in the  specification of the algorithm. They are perfect in the sense that they do not represent any hardware-level implementation and hence do not suffer from any noise.  Similarly, logical-level gate operations are also perfect.  They occur at the algorithmic level of specification. 
Until this point,  qubits always referred to such logical qubits. The logical level is the level of abstraction most suitable for high-level programming.

\emph{Physical qubits} are actual noisy qubits that decohere and lose their quantum properties after a short time -- too short for most quantum programs. To extend their lifetime, a large number of physical qubits can be combined with an error correction scheme to represent one logical qubit.

\emph{Hardware qubits} are the physical qubits constrained to a specific quantum device chip.  They each have a unique address.

During the logical layout step, we map quantum variables to specific logical qubits, taking into account locality, available operations, timing, and parallelism. Locality of logical qubits is important as it may only be possible to apply gates between neighboring qubits. Hence, if two logical qubits are far apart from each other, one has to rely on expensive teleportation or swapping of logical qubits. 
In this step, timing must also be introduced for parallel code execution, since different operations may have different durations that have to be taken into account for parallel scheduling of operations.

Note that early quantum computers will not have enough qubits to perform quantum error correction. In this case, we treat physical as logical qubits and skip the next error correction step.

\subsection{Quantum error correction (QEC)}
Unlike classical computers, quantum computers suffer readily from noise on very short time scale and require substantial overhead in both additional qubits and quantum gates to maintain a fault-tolerant and reliable computation.
Quantum error correction protects a logical qubit through redundancy, which is similar in nature to a classical repetition code, where a bit of information is copied into several bits and the information is decoded by taking a majority vote over the copies.  If the error rate on a given bit is low enough, that is, below a threshold, then the information will be protected.

Quantum information cannot be cloned due to the No Cloning Theorem. However, quantum circuits exist to distribute the information of a logical qubit across many physical qubits and to extract information on the types of errors through measurement. QEC protocols make use of this fact and allow quantum information to be protected by the use of quantum encoding and decoding circuits.
Ultimately, in any software-level QEC protocol, a logical qubit is encoded into a number of physical qubits, where the number depends on the error rates of the physical quantum device and the protection capacity, called the distance, of the selected quantum error-correcting code.
To date, one of the best QEC codes is the so-called surface code, which can protect against error rates of up to $1\%$ \cite{fowler12surface}.  

A software architecture must handle the mapping from logical qubit to the underlying physical qubits' representation.  
Performing this mapping requires knowledge of the number of physical qubits required to represent the one (or more) logical qubit(s), as well as the constraints of the physical mapping. 
Each logical quantum gate operation must also be encoded; the mapping must replace each logical operation by the encoded fault-tolerant operation for the given code.
The overlaying of quantum error correction on the logical computation is a non-trivial layout and scheduling task and will require substantial optimization routines to minimize resource overheads.
Note that in encoding each logical qubit and logical gate, a substantial number of physical qubits and computation time is required.

A QEC code also requires a decoding procedure~\cite{fowler12blossom,edmonds65}.  While classically one can decode the repetition code very simply with a majority vote,  most quantum decoding algorithms require more substantial computation.
The process of decoding is in fact {\it classical} and occurs on a classical processor.

The choice of QEC protocol depends on the hardware qubit properties, the fidelity of the hardware's quantum gates, and the number of logical gates in the quantum algorithm.
Thus, it will be crucial to have a QEC library available in order to select the QEC protocol that minimizes the overall resources required to complete the given quantum algorithm.
It will also be important to have very fast feedback and control between the classical and quantum hardware.  The classical processor performing the decoding must be fast enough to maintain pace with the quantum computer and decode errors at speed, so as to prevent the logical qubits from becoming too noisy.

\subsection{Physical layout generator and optimizer}

At this level, the physical qubits are mapped to actual hardware qubits. This mapping may be straightforward if the hardware layout is built with a specific error correction scheme in mind. In addition, some of the gate operations might require adjustments.

\subsection{Mapping to hardware}

The final step is the mapping of physical gate operations to hardware control of the specific device. For example, a hardware backend for superconducting qubits may map a native $X$ gate to a function call of an arbitrary wave form generator which will apply a microwave pulse corresponding to an $X$ gate on this qubit.

\section{Backends}\label{sec:backends}
Our architecture was specifically designed to support a variety of different backends, not just hardware backends but also software backends, e.g., simulators, emulators, and resource counters.
\\\\\textbf{\textit{Hardware backends. }}
There are many different competing technologies for quantum computer, including superconducting circuits, topological qubits, ion traps, spin qubits, and photonic qubits, just to name a few. Each of these technologies will have a unique set of native gate operations, and additionally each device a unique control interface.
\\\\\textbf{\textit{Simulators.}}
Simulators  play an important role in the development of quantum computers. At the lowest level, a simulator can be used to simulate the operation of the native hardware gates or at a higher level of logical qubits. Simulators can  include noise for different hardware technologies, which makes it possible to study its effects on the performance of quantum algorithms, especially on near-term quantum computers that will not have logical qubits yet. The size of the complex valued wave function representing the state of a quantum computer with $N$ qubits  grows as $2^N$. Since a gate operation can be implemented by a sparse matrix-vector multiplication such a simulation of a quantum computer is a sequence of matrix vector multiplications.  With current supercomputers, around 40-50 qubits can be simulated. 

Some gate operations, such as swap gates, may be simulated more efficiently by simply switching the indices of the qubit ordering rather than changing the exponentially large wave function.
\\\\\textbf{\textit{Emulators.}}
More efficient simulation is possible by emulating quantum computers at a higher level of abstraction.
Libraries may provide a way of emulating the functions they implement, which the emulator will use instead of performing a sequence of many gate operations. In the following, we list a few key example:
\begin{itemize}
\item  Mathematical functions, such as the modular exponentiation required for Shor's algorithm, do not need to be simulated through a large number of reversible gate operations, but can be emulated more efficiently by simply calling classical mathematical operations on each of the computational basis states. 
\item Quantum Fourier transforms can be emulated similarly by performing a fast Fourier transform on the wave function.
\item Repeated application of the same gate sequence for a large number of times $n$ may be optimized by building up a dense matrix representation $U$ of the actions of these gates and then performing repeated squaring to calculate $U^n$. Whether this is more efficient depends on the tradeoff between doing $O(n)$ sparse matrix-vector operations or $O(\log n)$ dense matrix-matrix operations.
\item Similarly, a quantum phase estimation, which essentially computes eigenvalues and eigenvectors of a unitary matrix, may be emulated more efficiently by diagonalizing the matrix for cases where the number of qubits is small.

\item When encountering measurements, an emulator can record the full expected distribution instead of sampling from it by performing random choices, as it is done in simulation.
\end{itemize}

Finally, oracular algorithms where no concrete quantum implementation of the oracle exists can still be emulated by providing a classical emulation of the oracle.
\\\\\textbf{\textit{Resource counters.}}
A resource counter keeps track of the number of qubits and gates that a certain algorithm would use when executed on a specific quantum computer. This is important information for designing and improving quantum algorithms. It is not sufficient that the asymptotic scaling of a quantum algorithm is better than that of the best known classical algorithm, but also that the crossover point in absolute time is small enough \cite{Wecker14a}. Estimating resources is important to guide optimization efforts and to identify realistic applications of quantum computers.

\section{Optimization and auto-tuning of libraries}
\label{sec:autotuning}
In order to reduce the cost of quantum library functions, an auto-tuning approach can be used. This is similar to the classical case, where linear algebra packages such as ATLAS automatically determine optimal tuning parameters. This is achieved by compiling different code versions and choosing the one yielding the best runtime. 	Due to the immense cost of optimization and auto-tuning, these should be done at the level of library functions, which are pre-compiled. This allows for reasonable compilation times of quantum programs that make use of such highly-tuned library functions.

Yet, trivially applying this auto-tuning approach is not feasible, given the vast amount of possible parameters in need of tuning. This can be seen when considering the addition operation as an example: There is a much wider variety of different adders in quantum than there is in classical computing. Some are classical addition algorithms that have been generalized to the quantum domain, such as Cuccaro et al.'s Ripple-Carry addition~\cite{cuccaro04}. Others are purely quantum, such as Draper's addition in Fourier space~\cite{draper00}. Which adder to use on which architecture depends on much more than just the operation count and the type of gates: In-place algorithms are favorable in terms of space constraints, which are much more severe in quantum computing. Yet, uncomputing intermediate results becomes more time-consuming in that case. Furthermore, good trade-offs between uncomputing intermediate results early and requiring more qubits must be found, since those temporary results will have to be recomputed in order to clear ancilla registers. Recently, this issue has been addressed by a compiler called \textsc{Revs}~\cite{parent15} which is able to reduce space complexity by a factor of 4 for large functions such as e.g., SHA-2 when compared to Bennett's method~\cite{bennett73}. Such approaches can be used to generate code for low- and high-level mathematical functions for which no hand-optimized implementations are available.

\section{Demonstration implementations}\label{sec:impl}
We have implemented a simplified version of the software architecture described above in several languages for demonstration purposes. Our implementation in {\em Python} uses runtime dispatch of high-level functions to library implementations, and then to backend-specific compilation of gates into hardware gates and the dispatch to a simulator or actual quantum hardware. The second implementation in {\em C++} uses template meta programming for compile-time dispatch to the appropriate implementation for the backends (resource estimator, simulator, emulator hardware calls), combined with additional optimizations performed on gate sequences at runtime.

Carrying out the quantum compilation at runtime (e.g., in Python) or at compile-time by the host language compiler (e.g., in {\em C++}) are only two options. Other possible approaches are to perform the quantum compilation by post-processing an intermediate representation of the program produced by the host language compiler, for example by transforming the LLVM representation of a {\em C++} program or the MSIL output of a C\# or F\# compiler \cite{solidtalk}.

\section{Summary}\label{sec:summary}

With the rapid development of small-scale quantum computers, it is necessary to develop a matching software infrastructures that goes beyond the simulators that have been developed so far.
Our approach to a quantum software architecture begins with an embedded domain-specific language by representing quantum types and operations through types and functions existing in a classical host language.  This leverages the capabilities of the host language while allowing substantial user-driven development. It also underlines the roles of quantum computers, namely as special purpose accelerators for existing classical codes.  By using an eDSL, we do not have to port the classical functionality to a newly developed quantum language.

In the near term, a quantum software architecture will allow the control of small-scale quantum devices and enable the testing, design, and development of components on both the hardware and software side.  We will witness more rapid innovations through the use of a common platform among quantum theorists, engineers, and experimentalists.
We believe that an instantiation of our software methodology will enable demonstrations of quantum programs for a variety of different devices in the very near future.

A high-level programming environment with optimized quantum emulation backends is important for the development of quantum algorithms and programs for future quantum computers. This will allow to demonstrate their usefulness and motivate further development of quantum hardware and software. In future work, we look forward to developing the various components and seeing a revolution in computation through scalable quantum computers.

\begin{acknowledgments}
We thank Alan Geller, Alexandr Kosenkov, Martin Roetteler, Svyatoslav Savenko,  Dave Wecker, and Nathan Wiebe for useful discussions and we acknowledge support by the Swiss National Science Foundation, the Swiss National Competence Center for Research and hospitality by the Aspen Center for Physics, supported by NSF grant \#PHY-1066293.
\end{acknowledgments}

\bibliography{references}

\appendix
\section{Appendix}

The state of one qubit can be represented using a two-dimensional complex vector as follows
\[
	\alpha\Ket0+\beta\Ket1 =\alpha\left(\begin{matrix}1\\0\end{matrix}\right)+\beta\left(\begin{matrix}0\\1\end{matrix}\right)=\left(\begin{matrix}\alpha\\\beta\end{matrix}\right)\;,
\]
where $\Ket0$ and $\Ket1$ are the basis states of a two-dimensional complex vector space and $\alpha,\beta\in\mathbbm C$ are complex amplitudes satisfying the normalization condition, i.e. $|\alpha|^2+|\beta|^2=1$. This condition ensures that the total probability of measurement outcomes sums up to 1.

For two qubits, the basis states are all possible configurations of two classical bits, i.e. $\Ket{00},\Ket{01},\Ket{10},\Ket{11}$. A general two-qubit state can be written in vector notation as
\begin{align*}
	&\alpha\Ket{00}+\beta\Ket{01}+\gamma\Ket{10}+\delta\Ket{11} 
	\\&=\alpha \left(\begin{matrix}1\\0\\0\\0\end{matrix}\right)+\beta\left(\begin{matrix}0\\1\\0\\0\end{matrix}\right)+\gamma\left(\begin{matrix}0\\0\\1\\0\end{matrix}\right)+\delta\left(\begin{matrix}0\\0\\0\\1\end{matrix}\right)
	=\left(\begin{matrix}\alpha\\\beta\\\gamma\\\delta\end{matrix}\right)\;,
\end{align*}
where $\alpha,\beta,\gamma,\delta\in\mathbbm C$. More generally, the quantum state of $n$ qubits can be represented by a complex vector of length $2^n$.

A unitary quantum operation on $n$ qubits can be written as a matrix of dimension $2^n\times2^n$. A few common examples of one- and two-qubit gates can be found in Table \ref{tbl:gates}. 
\begin{figure}[!t]
	\centering
	\includegraphics[width=.5\linewidth]{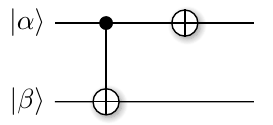}
	\caption{Quantum circuit representation of the example.}
	\label{fig:examplecirc}
\end{figure}
\\\\\textbf{\textit{Example.}}
Suppose two qubits are in single-qubit states $\Ket{\alpha}:=\alpha_0\Ket0+\alpha_1\Ket1$ and $\Ket{\beta}:=\beta_0\Ket0+\beta_1\Ket1$, the state of the entire system is the tensor product of the two individual states, which corresponds to the Kronecker product in vector notation, i.e.
\begin{equation}\label{eq:statevec}
	\left(\begin{matrix}\alpha_0\\\alpha_1\end{matrix}\right)\otimes\left(\begin{matrix}\beta_0\\\beta_1\end{matrix}\right)=\left(\begin{matrix}\alpha_0\beta_0\\\alpha_0\beta_1\\\alpha_1\beta_0\\\alpha_1\beta_1\end{matrix}\right)\;.
\end{equation}
Equivalently, this can be written as
\[
\alpha_0\beta_0\Ket{00}+\alpha_0\beta_1\Ket{01}+\alpha_1\beta_0\Ket{10}+\alpha_1\beta_1\Ket{11}\;.
\]
As a side note: Not all two-qubit states can be written as a product of single-qubit states. Such states are said to be entangled and play a crucial role in quantum computing.

The result of applying a CNOT gate to these two qubits can be obtained by a matrix-vector multiplication of the CNOT gate matrix in Table \ref{tbl:gates} with the state vector in \eqref{eq:statevec}:
\[
	\left(\begin{matrix}
	1 & 0 & 0 & 0\\
	0 & 1 & 0 & 0\\
	0 & 0 & 0 & 1\\
	0 & 0 & 1 & 0
	\end{matrix}\right)\cdot\left(\begin{matrix}\alpha_0\beta_0\\\alpha_0\beta_1\\\alpha_1\beta_0\\\alpha_1\beta_1\end{matrix}\right)=\left(\begin{matrix}\alpha_0\beta_0\\\alpha_0\beta_1\\\alpha_1\beta_1\\\alpha_1\beta_0\end{matrix}\right)
\]
Next, we want to apply a NOT gate to the first qubit and leave the second one unchanged. This can be accomplished by multiplying the state vector with the matrix
\[
	X\otimes\mathbbm 1_2=\left(\begin{matrix}0 & 1\\1 & 0\end{matrix}\right)\otimes\left(\begin{matrix}1 & 0\\0 & 1\end{matrix}\right)=\left(\begin{matrix}
	0 & 0 & 1 & 0\\
	0 & 0 & 0 & 1\\
	1 & 0 & 0 & 0\\
	0 & 1 & 0 & 0
	\end{matrix}\right)\;,
\]
which yields
\[
	\left(\begin{matrix}
		0 & 0 & 1 & 0\\
		0 & 0 & 0 & 1\\
		1 & 0 & 0 & 0\\
		0 & 1 & 0 & 0
		\end{matrix}\right)\cdot\left(\begin{matrix}\alpha_0\beta_0\\\alpha_0\beta_1\\\alpha_1\beta_1\\\alpha_1\beta_0\end{matrix}\right)=\left(\begin{matrix}\alpha_1\beta_1\\\alpha_1\beta_0\\\alpha_0\beta_0\\\alpha_0\beta_1\end{matrix}\right)\;.
\]
The corresponding circuit diagram is depicted in Figure~\ref{fig:examplecirc}.

\begin{table}
	\centering
	\vspace{20pt}
	\renewcommand{\arraystretch}{1}
	\begin{tabular}{>{\centering}m{2cm} >{\centering}m{2cm} >{\centering\arraybackslash}m{2cm}}
	\toprule
	Gate & Matrix & Symbol\\\hline
 NOT or X gate& \(\displaystyle\left(\begin{matrix}0 & 1\\1 & 0\end{matrix}\right)\)&\includegraphics[width=2cm]{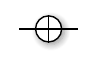}\\
 Y gate & \(\displaystyle\left(\begin{matrix}
 			0 & -i\\
 			i & 0
 			\end{matrix}\right)\)&\includegraphics[width=2cm]{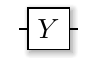}\\
	Z gate & \(\displaystyle\left(\begin{matrix}
			1 & 0\\
			0 &-1
			\end{matrix}\right)\)&\includegraphics[width=2cm]{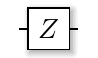}\\
	Hadamard gate & \(\displaystyle\frac 1{\sqrt2}\left(\begin{matrix}
			1 & 1\\
			1 &-1
			\end{matrix}\right)\)&\includegraphics[width=2cm]{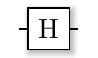}\\
	S gate& \(\displaystyle\left(\begin{matrix}
			1 & 0\\
			0 & i
			\end{matrix}\right)\)&\includegraphics[width=2cm]{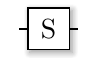}\\
	T gate & \(\displaystyle\left(\begin{matrix}
			1 & 0\\
			0 & e^{i\pi/4}
			\end{matrix}\right)\)&\includegraphics[width=2cm]{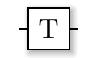}\\
	Rotation-Z gate & \(\displaystyle\left(\begin{matrix}
			e^{-i\theta/2} & 0\\
			0 & e^{i\theta/2}
			\end{matrix}\right)\)&{\centering\includegraphics[width=2.5cm]{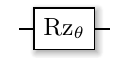}}\\
	Controlled NOT (CNOT) & \(\displaystyle\left(\begin{matrix}
			1 & 0 & 0 & 0\\
			0 & 1 & 0 & 0\\
			0 & 0 & 0 & 1\\
			0 & 0 & 1 & 0
			\end{matrix}\right)\)&\includegraphics[width=2cm]{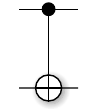}\\\bottomrule
	\end{tabular}
	\caption{Standard quantum gates with their corresponding matrices and symbols.}
	\label{tbl:gates}
\end{table}

\end{document}